\documentclass[%
reprint,
 amsmath,amssymb,
aps,
]{revtex4-2}

\usepackage{graphicx}
\usepackage{dcolumn}
\usepackage{bm}

\usepackage{enumitem} 
\usepackage{amsmath}
\usepackage{color}
\usepackage{soul}

\def\e{\begin{equation}}
\def\f{\end{equation}}
\def\_#1{{\bf #1}}

\def\.{\cdot}

\def\@#1{_{\rm #1}}

\begin{document}


\title{Effective Mid-Range Wireless Power Transfer with Compensated Radiation Loss}
\author{N.~Ha-Van}%
 \email{nam.havan@aalto.fi}
\author{C.~R.~Simovski}
\author{F.~S.~Cuesta}%
\author{P.~Jayathurathnage}%
\author{S.~A.~Tretyakov}%
\affiliation{%
 Department of Electronics and Nanoengineering, Aalto University, P.O. Box 15500, Aalto FI-00076, Finland
}%


\begin{abstract}
In conventional inductive wireless power devices, the energy is transferred via only reactive near fields, which is equivalent to non-radiative F\"orster energy transfer in optics. Radiation from transmitting and receiving coils is usually considered as a parasitic effect that reduces the power transfer efficiency. As long as the distance between the two antennas is small as compared to the antenna size, conventional WPT devices offer rather high power transfer efficiency, of the order of 80-90\%. However, for larger distances, the transfer efficiency dramatically drops, making such devices not practical. In this paper, we develop a dynamic theory of wireless power transfer between two small loop antennas, clarify the role of far-field radiation, and find a possibility to realize efficient wireless power transfer at large distances utilizing the regime of radiation suppression due to optimized mutual  dynamic interactions between the transmitting and receiving antennas. The analytical results have been validated by simulations and measurements, and they open a possibility to greatly expand the range of distances of compact wireless power transfer devices. The developed theory can be applied also to coupling between antennas of different types and to energy transfer between nano-objects. 
\end{abstract}

\maketitle


\section{Introduction}\label{sec:Introduction}

Wireless power transfer (WPT) technologies become more and more important for diverse applications, such as charging of mobile telecommunication devices, electric vehicles, implantable medical devices, robots, wearable electronics, and in energy harvesting systems (see e.g. in  \cite{KursScience,RonHui_review,Brown1984}). During recent decades, especially proliferation of wireless devices has motivated fast developments of wireless powering and charging technologies. 

All WPT systems can be classified into near-field and far-field ones (see e.g. in \cite{JenshanLin2013}). 
When the distance between the transmitting and receiving antennas is small compared to the wavelength, the reactive near fields at the receiver position are much stronger than slower decaying fields of propagating waves created by the transmitting antenna. In this short-range WPT regime, the power is transferred predominantly by the near fields, whereas the radiation is usually considered as a parasitic factor resulting in some radiation loss that decreases  efficiency (see e.g. in \cite{Siqi2014,Shamsul2022,2014_Hui_2D-3D,Nam2022,Lee_Fred_Load_Detection,Agarwal2017,Mashhadi2018,Nam_biomedical}). Respectively, it appears that in this regime it is desirable to suppress antenna radiation. However, the radiative fields are present also in the near zone and they may contribute to the received power. 

If the distance between the transmitting and receiving antennas is electrically large, the near fields of the transmitting antenna become negligibly small as compared with the radiation fields. For electrically small and weakly directive antennas, the transfer efficiency becomes very small. For this reason, in long-range radiative WPT systems  high-directivity antennas are used in both  receiving (Rx) and transmitting  (Tx) devices. Obviously, in this case, radiation is used as the main power-transfer mechanism, and radiation losses can be reduced only by using higher-gain antennas of large sizes in comparison with the wavelength.

In this paper, we consider the intermediate case of mid-range wireless power transfer, where at the receiver position near fields of the transmitter significantly decay and the slower-decaying  radiative fields become comparably strong or even stronger. In this situation, radiation of energy into far zone is still unwanted, as this is one of the loss mechanisms. On the other hand, radiative fields can significantly contribute to the power transfer to the receiving antenna.  
In practice, this situation corresponds to scenarios where the distance between  the transmitting and receiving antennas is large as compared to the sizes of the antennas, while the antennas themselves are electrically small. Realization of such WPT  systems would allow wireless power transfer using compact devices, as the transfer distance can be large as compared to both connected devices. It is expected that the power transfer efficiency, in this case, will be lower than for near-field coupling devices working at small transfer distances, but it may be possible to realize reasonable efficiency also at large separations by exploiting coupling by both near fields  and radiative fields. 

In this study, we develop the dynamic theory of wireless power transfer between two magnetic dipole antennas and use it to study the role of radiation fields (intermediate and far-zone) in transfer of power. We  answer the following questions: What is the dominant role of dynamic (far-zone) fields --- disadvantage of parasitic radiation loss or an  advantage of higher electromotive force induced in the receiving antenna? Is it possible to create WPT devices where radiation into far zone is suppressed while the dynamic fields between the antennas effectively deliver power to the load?   How to properly engineer the optimal regime of long-distance WPT between small antennas?  
To our knowledge, these questions have not been answered in the available literature. In spite of the great diversity of radio-frequency WPT systems, the comparative role of the radiative and non-radiative regimes was not properly  elucidated. 
An important question is the suppression of radiation in the short-range systems of non-radiative WPT. Dynamic coupling between two loop antennas was recently considered in \cite{Shamonina}, where this question was addressed, focusing on visualizations of the power flow in space. In that paper, it has been found that it is possible to maximize the ratio of the delivered power to the power available from the source and at the same time minimize the radiated power by optimizing the load resistance for a fixed value of the internal resistance of the source.

In this paper, we find a simple formula for the optimal load impedance and the corresponding maximized power transfer efficiency that are valid for arbitrary internal impedance of the source. 
We use these results to understand physical mechanisms of power delivery accounting for dynamic interactions, clarify the role of far fields, and find the optimal frequency range where the parasitic radiation into the far zone is effectively suppressed while dynamic, far-zone fields strongly contribute to the power delivered to the source. 

The developed  theory of the optimal load impedance and the maximized power transfer efficiency can be used for optimizing coupling between arbitrary antennas, including electric dipoles. However, we focus the study on loop antennas, because electric dipoles are used in WPT devices quite rarely, mainly in very short-range systems for high power transfer via capacitive coupling \cite{Qu2015,Tran2019} and in WPT systems operating in lossy media, e.g., for biomedical implants \cite{abdulfattah2019performance}.  
Most  WPT systems operating in free space utilize MDs implemented as coils \cite{KursScience,RonHui_review,Tse2019converterforcharging,Zhang2020,Wang2021}.  

This study is also relevant for understanding and use of  non-radiating antennas -- those creating only near fields, such as anapole antennas \cite{Anapole2015, Anapole2019}. The evident advantage of  anapole antennas is negligibly small parasitic radiation for all arrangements of the two antennas where mutual coupling of the antennas does not destroy the anapole properties. 
In paper \cite{Zanganeh2021}, anapole antennas were claimed to be more beneficial for the short-range WPT than MDs due to suppressed radiation. The use of the theory developed in this work allows  us  to understand full implications of radiation cancellation for power coupling between antennas. 

The paper is organized as follows. In Section~\ref{sec:WPT_Analysis}, we theoretically analyze the dynamic coupling between two small antennas WPT and elaborate the model of power transfer efficiency ($PTE$)  applicable for a maximally broad frequency range and maximal range of distances. In Section~\ref{sec:Loop_Analysis}, we study two key arrangements of Tx and Rx loops; coaxial and coplanar. Finally, the developed analysis is verified by full-wave simulations and experiments in Section~\ref{sec:Experiment}. 

\begin{figure}
	\centering
	\includegraphics[width=1\linewidth]{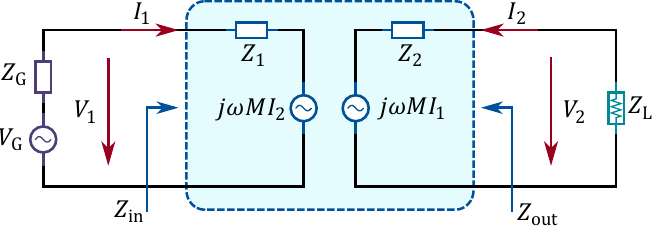}
	\caption{Equivalent circuit model of the WPT system.} 
	\label{fig:EquiCir}
\end{figure}

\section{WPT System Analysis}\label{sec:WPT_Analysis}

Figure~\ref{fig:EquiCir} shows the equivalent circuit of WPT systems formed by two coupled antennas. The  two antennas are represented by their individual input impedances $Z_1$ and $Z_2$. The transmitting antenna is fed by a voltage source $V_{\rm G}$ with internal impedance $Z_{\rm G}$, and the receiving antenna is loaded by the load impedance $Z_{\rm L}=R_{\rm L}+jX_{\rm L}$, whose real part $R_{\rm L}$ is the useful load.

Similarly to  \cite{Shamonina}, we find it convenient to express 
the electromotive forces (EMFs) induced by antennas in one another through the complex coefficient $M=M_{\rm R}+jM_{\rm I}$ instead of the commonly used mutual impedance,  generalizing the notion of mutual inductance to dynamic regimes, applicable to arbitrary distances between the antennas:
\begin{align}
    V_{\rm emf,\it i}=j\omega MI_i . \label{eq:1}
\end{align}
Here, $i=1,2$ mark the first and second antennas. For inductive wireless power transfer devices and in the quasi-static limit, $M $ is real-valued and equals to the usual mutual inductance between the two coils. The value  $Z_m=j\omega M$ is the mutual impedance, not to be mixed with the induced impedances $Z_{i1}=V_{\rm emf,1}/I_1=Z_mI_2/I_1$ and $Z_{i2}=V_{\rm emf,2}/I_2=Z_mI_1/I_2$). 
For voltages $V_{1,2}$ we have 
\begin{align}
    V_1 = Z_1 I_1 + j\omega M I_2 = V_{\rm G} - I_1 Z_{\rm G} \label{eq:2} \\
    V_2 = I_2 Z_{\rm L}= -I_2 Z_2 + j\omega M I_1  , \label{eq:3}
\end{align}
where $Z_{1,2}=R_{1,2}+jX_{1,2}$ are complex-valued antenna input impedances. The impedances seen from the source and the load (see Fig.~\ref{fig:EquiCir}) can be expressed in terms of the complex-valued mutual inductance $M$ in the usual form:  
\begin{align}
    Z_{\rm in} =\frac{V_1}{I_1} = Z_1 + \frac{\omega^2 M^2}{Z_2 +Z_{\rm L}}  \label{eq:4} \\
    Z_{\rm out} = \left.\frac{V_2}{I_2}\right|_{V_{\rm G}=0} = Z_2 + \frac{\omega^2 M^2}{Z_1 +Z_{\rm G}} .\label{eq:5}
\end{align}

Aiming to the maximal power coupling between antennas, we use the conventional definition of the power transfer  efficiency $PTE$ as the ratio of the power delivered to the load $P_{\rm L} $ and the power accepted by the input port of the transmitting antenna  $P_{\rm in}$ 
(see e.g. in \cite{RonHui_review}):  
\begin{align}
    {PTE} \equiv \frac{P_{\rm L}}{P_{\rm in}} = \frac{|I_2|^2 R_{\rm L}}{\mathfrak{Re}\{V_1 I_1^*\}}.   \label{eq:6} 
\end{align}
In this definition, the transfer efficiency does not depend on the internal impedance of the source $Z_{\rm G}$, because the received power is normalized to the power accepted by the transmitting antenna. 
Using (\ref{eq:3}) and (\ref{eq:4}), we write Eq. (\ref{eq:6}) in the form
\begin{align}
    {PTE} = \frac{|I_2|^2 R_{\rm L}}{ |I_1|^2 \mathfrak{Re}\{Z_{\rm in}\}}=\left|\frac{\omega M}{Z_2 + Z_{\rm L}} \right|^2 \frac{R_{\rm L}}{\mathfrak{Re}\{Z_{\rm in}\}}.  \label{eq:7} 
\end{align}

The optimal load $Z_{\rm L,opt}$ maximizing the $PTE$ can be found by nullifying the derivatives~\cite{Prasad_two_port}: 
\begin{equation}
\frac{\partial {PTE}}{\partial X_{\rm L}} = 0,\quad  \frac{\left.\partial {PTE}\right|_{X_{\rm L,opt}}}{\partial R_{\rm L}} = 0, 
\end{equation}
from which we obtain the reactance and resistance of the optimal load that depends on the operational frequency and complex-valued mutual inductance:
\begin{equation}
\begin{aligned}
    & X_{\rm L,opt} = - X_2 -\frac{\omega^2 M_{\rm R} M_{\rm I}}{R_1}, \label{eq:8}
\end{aligned}
\end{equation}  
\begin{equation}
\begin{aligned}
    & R_{\rm L,opt} = \frac{\sqrt{R_1 R_2 - \omega^2 M_{\rm I}^2}\sqrt{R_1 R_2 + \omega^2 M_{\rm R}^2}}{R_1}. \label{eq:9}  
\end{aligned}
\end{equation}
This result generalizes expressions in  \cite{Prasad_two_port} taking into account dynamic interactions between Tx and Rx antennas. 
Substituting~(\ref{eq:8}) and~(\ref{eq:9}) into~(\ref{eq:7}), the  $PTE$ of devices loaded by optimal loads after some algebra can be expressed in a very simple form: 
\begin{align}
PTE=1-\dfrac{2}{1+\sqrt{\dfrac{R_1R_2+\omega^2M_{\rm R}^2}{R_1R_2-\omega^2M_{\rm I}^2}}}. 
\label{eq:10} 
\end{align}
Parameters $M_{\rm R},\, M_{\rm I}, \, R_1,\, R_2$ can be found analytically, numerically, or experimentally.

The same expression for $PTE$ can also be written in form of the classical definition of impedance parameters, where $R_1=r_{11}$, $R_2=r_{22}$, $\omega M_{\rm I}=r_{12}$, and $\omega M_{\rm R}=x_{12}$ are the real and imaginary parts of the corresponding  components of the impedance matrix of the wireless link:
\begin{align}
PTE=1-\dfrac{2}{1+\sqrt{\dfrac{r_{11}r_{22}+x_{12}^2}{r_{11}r_{22}-r_{12}^2}}}. 
\label{eq:11} 
\end{align}
In this form, this result can be used for optimization of power transfer between arbitrary emitters and receptors, not necessarily loop antennas.

The found  simple analytical expression for the power transfer efficiency at the optimal load allows us to perform a broadband analysis of the efficiency,  find the optimal operational frequency and clarify the role of radiation in short- and mid-range WPT systems. Before going to the analysis of systems with specific antennas, let us make some general observations. First, it is obvious that in order to increase $PTE$ we should increase the value of the square root in \eqref{eq:10} or \eqref{eq:11}. Assuming that the two resistances are fixed, it is clearly beneficial to increase $M_{\rm R}$, which, in the case of electrically small distances, means simply that the transfer efficiency is larger at smaller transfer distances. More interestingly, we see that increasing the imaginary part of $M$ also leads to higher efficiency. The denominator under square root cannot reach zero (which would mean 100\% efficiency), because the mutual impedance cannot be higher than the impedance of both coupled circuits, but it can become rather  small under some conditions. In the following, we will discuss these two factors at an example of two coupled electrically small loops.

Since the power transfer efficiency \eqref{eq:6} and the optimal load impedance (\ref{eq:8})--(\ref{eq:9}) do not depend on the internal impedance of the source $Z_{\rm G}$, that can be chosen based on application requirements. If the WPT device should deliver the maximal power to the load, the source impedance can be conjugate matched to $Z_{\rm in}$. In most applications, however,  it is preferable to maximize the end-to-end power transfer efficiency, in which case $Z_{\rm G} $ is made as small as possible, to minimize dissipative losses. Importantly, our conclusions regarding optimization of $PTE$ remain valid for any value of $Z_{\rm G}$ as long as the load impedance $Z_{\rm L} $ is equal to the optimal load given by (\ref{eq:8}) and~(\ref{eq:9}). If the load resistance is different, a matching circuit can be used at the receiver side. 
This approach and the ultra-broad range of possible operation frequencies is qualitatively different from that of \cite{Shamonina}, where the optimization was done for the case when $R_{\rm G}\equiv{\rm Re}(Z_{\rm G})=50~\Omega$, and the operation frequency range was specified in advance.  

\begin{figure*}
	\centering
	\includegraphics[width=0.7\textwidth]{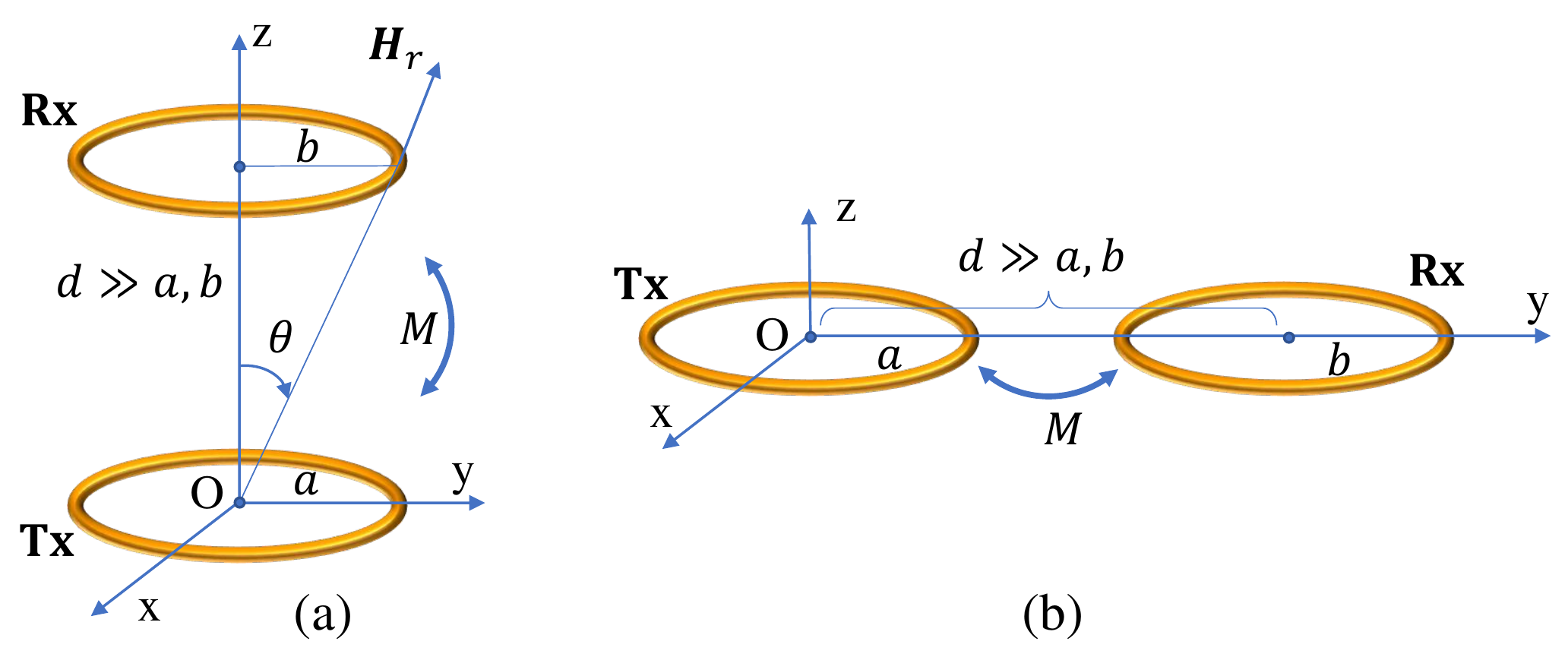}
	\caption{Two magnetic dipoles in two case studies: a) coaxial position; b) coplanar position. 
	} 
	\label{fig:Position}
\end{figure*}

\section{Analysis of a System Based on Loop Antennas}\label{sec:Loop_Analysis}

Our goal is to clarify the role of intermediate and far fields  in WPT systems in the maximally broad range of frequencies and for both short and middle ranges of distances, that is, when the distance between antennas can be comparable with the wavelength or electrically small. With this goal in mind,  we restrict the size of both antennas so that in the whole possible operational frequency range the antennas remain electrically small.  Otherwise, the comparison of the two regimes would be veiled by the size resonances of antennas.

In order to clarify the role of radiative fields, we consider small loop antennas and study WPT for coaxial and coplanar arrangements of the Tx and Rx antennas, as shown in Fig.~\ref{fig:Position}. Because the loops are electrically small, we model them as magnetic dipoles (MD). In the coaxial arrangement, the loops are coupled solely by near fields, since a magnetic dipole  does not radiate along its magnetic moment. In this case, there is no radiative WPT, and the radiation resistances of the loops are purely parasitic contributions to resistances  $R_{1,2}$. Thus, in order to increase  $PTE$ it is desirable to realize a regime of suppressed radiation. In the coplanar arrangement, the  situation is nearly opposite. 
In this case, the quasi-static mutual coupling is weak because the mutual inductance is small, and the radiative WPT can dominate already at mid-range distances. 

\subsection{Analytical model}
\subsubsection{Coaxial Arrangement}
Let us consider two loops, Tx and Rx, with the radii $a$ and $b$, respectively, positioned in free space at  distance $d$ between them, as it is shown in Fig.~\ref{fig:Position}(a). In the spherical coordinate system, the magnetic field 
components are as follows (see e.g. in \cite{Balanis}):
\begin{align}
    H_r =j\frac{ka^2I_1\cos{\theta}}{2d^2} \left[ 1+\frac{1}{jkd}\right]e^{-jkd}, \label{eq:12} \\
    H_\theta =-\frac{(ka)^2I_1\sin{\theta}}{4d} \left[ 1+\frac{1}{jkd}-\frac{1}{(kd)^2}\right]e^{-jkd}. \label{eq:13}
\end{align}
Here $k=\omega/c$ is the wave number in free space. 
When $a,b\ll d$, we can approximate $H_z \approx \rm const \it(x,y)$ in the area $z=d, x^2+y^2<b^2$, i.e., the $z$-component of the magnetic field created by Tx in the area of Rx loop is nearly equal to its $r$-component at the receiving loop center:
\begin{align}
    H_z \approx H_r  =  
     j\frac{k A_{\rm T} I_1}{2\pi d^2} \left[ 1+\frac{1}{jkd}\right]e^{-jkd},
    \label{eq:14}
\end{align}
where $A_{\rm T}=\pi a^2$. From (\ref{eq:14}) and Faraday's law $V_{\rm emf,2}=-j\omega \mu_0 \pi b^2H_z$, we find the complex mutual inductance $M$ defined by Eq.~(\ref{eq:1}) as 
\begin{align}
M \equiv M_{\rm R}+jM_{\rm I}= \frac{j\mu_0 k A_{\rm T}A_{\rm R}}{2\pi d^2} \left[ 1+\frac{1}{jkd} \right] e^{-jkd} \label{eq:15},
 \end{align}
where $A_{\rm R}=\pi b^2$ is the area of the Rx loop. In the quasi-static limit $kd\ll 1$, and formula~(\ref{eq:15}) transits to the known mutual inductance of two coaxial loops, applicable when $d\gg \max(a,b)$ \cite{Grover}. 

\subsubsection{Coplanar Arrangement}
Now we consider two loops positioned at the same $xy$-plane, as shown in Fig.~\ref{fig:Position}(b). If $a,b\ll d$, we can write  \begin{align}
    H_{z} \approx -\frac{k^2 A_{\rm T} I_1}{4\pi d} \left[ 1+\frac{1}{jkd}-\frac{1}{(kd)^2}\right]e^{-jkd}. 
    \label{eq:16}
\end{align}
Then we have for $M$ the expression: 
{\small
\begin{align}
M \equiv M_{\rm R}+jM_{\rm I}= -\frac{\mu_0k^2A_{\rm T}A_{\rm R}}{4\pi d} \left[ 1+\frac{1}{jkd}-\frac{1}{(kd)^2} \right] e^{-jkd}. \label{eq:17}
\end{align}
}
In the low-frequency limit, this expression also transits to the known quasi-static mutual inductance, when $d\gg \max(a,b)$ \cite{Grover}.  
When $a=b$, the quasi-static approximation is applicable with high accuracy when $d>4a$.

\subsubsection{Input Resistances of Tx and Rx loops}

The input resistances $R_1$ and $R_2$ of the Tx and Rx antennas  are the  sums of the Ohmic resistances $R_{\rm O1,O2}$ and radiation resistances
$R_{\rm r1,r2}$, i.e., $R_{1,2}=R_{\rm r1,r2}+R_{\rm O1,O2}$. For electrically small loops made of thin round wires in the regime of strong skin effect, we have (e.g., \cite{Balanis}):
\begin{align}
R_{\rm O1} = \frac{a}{r_0}\sqrt{\frac{\omega \mu_0}{2\sigma_w}}, \quad R_{\rm O2} = \frac{b}{r_0}\sqrt{\frac{\omega \mu_0}{2\sigma_w}}. \label{eq:18} \\
R_{\rm r1} = \eta \frac{\pi}{6} (ka)^4, \quad  R_{\rm r2} = \eta  \frac{\pi}{6} (kb)^4. \label{eq:19}
\end{align}
Here, $r_0$ is the radius of the antenna wire, $\sigma_w$ is the conductivity of the wire, and $\eta=\sqrt{\mu_0/\varepsilon_0}$ is the free-space impedance. 
In numerical examples, we assume copper wires with  $\sigma_w=58.7\times10^6$~S/m, $a=b=36$~mm, and $r_0=2$~mm.

Using formulas~(\ref{eq:15}),~(\ref{eq:17}), and~(\ref{eq:19}),  it is easy to verify that $\omega^2M_{\rm I}^2\le R_{\rm r1}R_{\rm r2}$ for $d\ge \max(a,b)$ in both coaxial and coplanar cases, confirming that in Eq.~(\ref{eq:10})
the expression under the square root is always positive.  

\begin{figure}[t]
	\centering
	\includegraphics[width=0.7\linewidth]{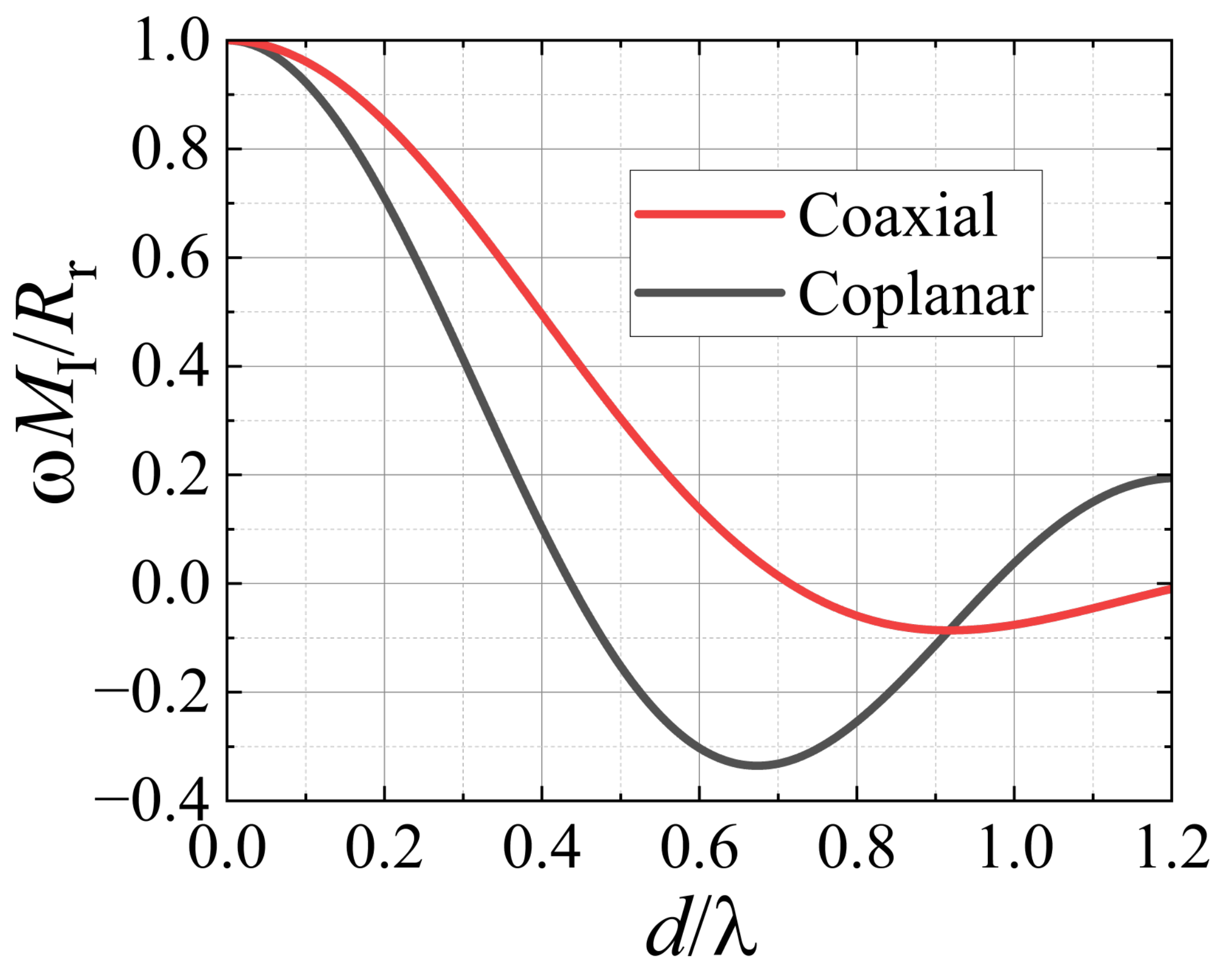}
	\caption{Ratio of the mutual resistance to the radiation resistance for coaxial and coplanar arrangements of the Tx and Rx loops versus the normalized distance $d/\lambda$ between the loop centers. 
	} 
	\label{fig:kappa}
\end{figure}

\begin{figure*}[t]
	\centering
	\includegraphics[width=\textwidth]{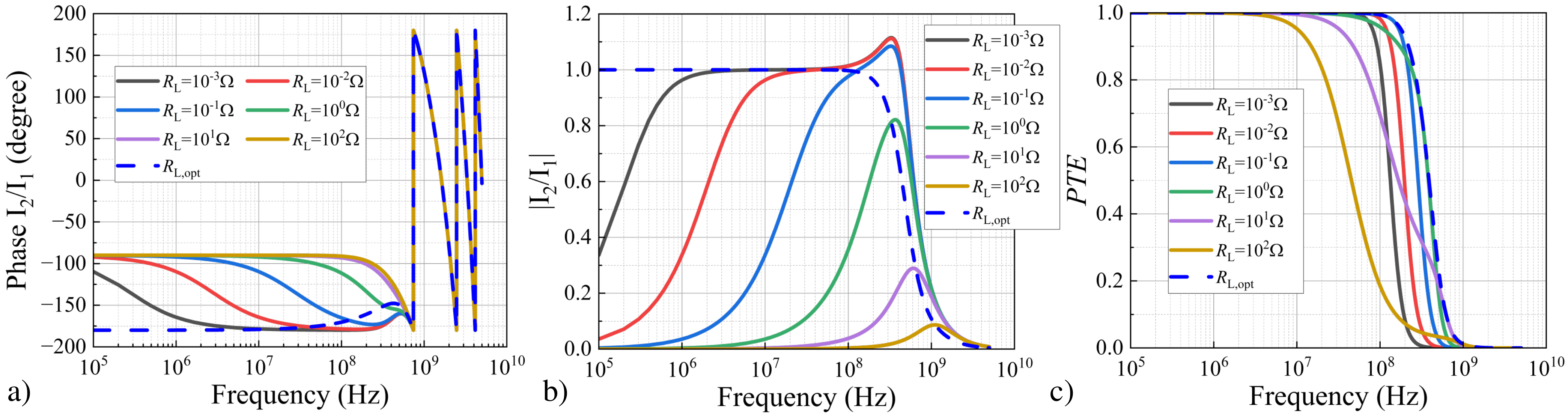}
	\caption{Mid-range coupling between two lossless loops: a)  phase difference between the loop currents; b) magnitude of the currents ratio $I_2/I_1$; c) $PTE$ for different values of the load resistance $R_{\rm L}$. The dashed curves correspond to the optimal resistance, maximizing $PTE$.}
	\label{fig:rad_lossless}
\end{figure*}

\subsection{Suppression of Radiation of Power into Space}

As discussed in Sec.~\ref{sec:Introduction}, it is important to clarify the role of radiation fields, since on one side they create parasitic radiation of power into space, but on the other hand, far fields can carry power to the receiver. In this part, we consider this question for coaxial loops. 

Let us first discuss the case when the dissipation of energy in both transmitting and receiving coils can be neglected. Then, the quasi-static theory of wireless power transfer between two loops tells that $PTE=1$ identically, because the power accepted by the transmitting antenna can go only to the load in the receiver. However, it is expected that the fully dynamic model will show finite transfer efficiency even in the limit of zero distance between antennas, because the antennas will radiate some power into space. Thus, in \cite{Shamonina} it was claimed that suppression of radiation into far zone is a useful mean to increase power transfer efficiency. Let us analyse this issue using the analytical formulas for the complex-valued mutual inductance and power transfer efficiency. 
First, we note that in the absence of losses in coils, the resistances $R_{1,2}$ in \eqref{eq:10} contain only radiation resistances of the two loops \eqref{eq:19}. On the other hand, the term $\omega M_{\rm I}$ in \eqref{eq:10} is the mutual resistance of the two antennas. It is obvious that in the limit $d\rightarrow 0$, the mutual resistance becomes equal to the radiation resistance of the loops (for simplicity we consider two identical loops), and the $PTE$ \eqref{eq:10} tends to unity, although the radiation of energy from both loops is fully accounted for. Physically, this means that at small transfer distances, selection of the optimal load value given by \eqref{eq:8}--\eqref{eq:9} leads to a WPT system with suppressed radiation into far zone. 

Perfect radiation suppression takes place only in the limit of zero distance, because the mutual resistance decays with increasing transfer distance. For the case of two coupled loops, this decay is illustrated in Fig.~\ref{fig:kappa}. 
We see that for $d<0.2\lambda$ the mutual resistance exceeds 80\% of the radiation resistance. This means that if the currents in the loops have opposite phases, the dynamic interaction of the two loops will cancel at least 80\% of radiation into space, enhancing the received power. 
To show that the mutual coupling namely decreases the radiation, we derive the ratio between $I_2$ and $I_1$ for the case when $R_{\rm L}=R_{\rm L,opt}$ and $X_{\rm L}=X_{\rm L,opt}$, which reads 

{\small
\begin{align}
\dfrac{I_2}{I_1}=\frac{-R_1 \omega (M_{\rm I}+ j M_{\rm R})}{R_1 R_2 + j \omega^2 M_{\rm R}M_{\rm I}+\sqrt{(R_1 R_2 - \omega^2 M_{\rm I}^2)(R_1 R_2 + \omega^2 M_{\rm R}^2)}}.
\label{eq:i_ratio}
\end{align}
}

It can be observed from \eqref{eq:i_ratio} that when $R_1=R_2=\omega M_{\rm I}$, the ratio between the currents indeed becomes equal to $-1$, which indicates that the currents have equal amplitudes and opposite phases ($180 ^{\circ}$). Obviously, the magnetic dipole radiation of the set of two antennas is suppressed in this case. 
Next, we plot the ratio of the currents in the two loops, $I_2/I_1$, both the amplitude and phase, see Figs.~\ref{fig:rad_lossless}(a) and (b). We take as an example the case when the distance between the loops is five times larger than the loop radius, $d=5a$. We see that at the optimal load (dashed curves) the regime of radiation suppression holds in the low frequency range, since the currents in the two loops are approximately equal in the amplitude and opposite in phase.   

Thus, such set of two antennas radiates into far zone only via its higher-order multipoles. An estimation of the ratio between the radiated power from a single magnetic dipole and the power radiated from a quadrupole formed by two counter-directed dipoles of the same amplitude (both in the coaxial and coplanar arrangements) gives 
\begin{equation}
\frac{P_q}{P_d}=\frac{1}{48\pi }(kd)^2,
\end{equation}
which is approximately $6\cdot 10^{-4}$ for $kd=0.5$. Thus, for the case of mid-range WPT distances, this higher-order multipole radiation can be neglected. 

Corresponding calculations for non-optimal values of the load resistance (solid curves in Fig.~\ref{fig:rad_lossless}) show that radiation is compensated in a certain frequency range but not in  the low-frequency regime. However, the analysis of the $PTE $ dependence [Fig.~\ref{fig:rad_lossless}(c)] reveals that the efficiency remains practically unity in a broad frequency range, also at low frequencies even for non-optimal loads, which is due to the fact that radiation is in any case weak at low frequencies. The frequency of the sharp drop of efficiency at high frequencies corresponds approximately to the frequencies above the first null of the dependence of the mutual resistance on the frequency (Fig.~\ref{fig:kappa}). The value $d/\lambda=0.5$ corresponds to approximately $8\cdot 10^8$~Hz for the chosen example value of $d$. We conclude that if the dissipative losses in the loop antennas can be neglected, the radiation into far zone is effectively suppressed due to dynamic antenna interactions up to the transfer distances of the order of one wavelength. Obviously, there is no need to use any special means to suppress parasitic radiation losses in this regime.

\begin{figure*}[t]
	\centering
	\includegraphics[width=\textwidth]{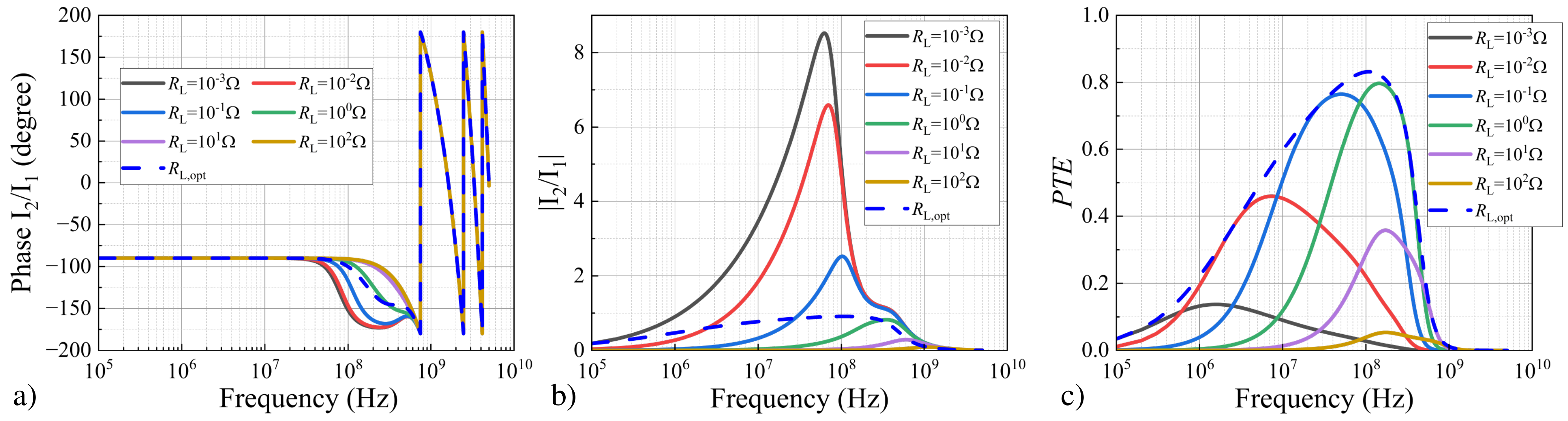}
	\caption{Mid-range coupling between two lossy loops: a)  phase difference between the loop currents; b) magnitude of the currents ratio $I_2/I_1$; c) $PTE$ for different values of the load resistance $R_{\rm L}$. The dashed curves correspond to the optimal resistance, maximizing $PTE$.
	} 
	\label{fig:rad_lossy}
\end{figure*}

Let us next consider mid-range interactions of loop antennas taking onto account dissipation in the loops. Figure~\ref{fig:rad_lossy} shows the same plots as Fig.~\ref{fig:rad_lossless}, but accounting for dissipative losses in both loops, as defined by Eq.~\eqref{eq:18}. We see that in this case selecting the optimal load impedance we realize the regime of effective suppression of radiation, but only in a certain frequency range, approximately between $10^8$ and $10^9$~Hz. In this range, the currents in the two loops are approximately of the same amplitude and of the opposite phase. At low frequencies, the phase difference between the currents is rather close to $90^\circ$, for a broad range of the load resistances (in all cases, the load reactance is the optimal one). This conclusion follows from \eqref{eq:i_ratio} by setting the imaginary part of the mutual inductance $M_{\rm I}$ to zero.  

\begin{figure}[!h]
	\centering
	\includegraphics[width=0.75\linewidth]{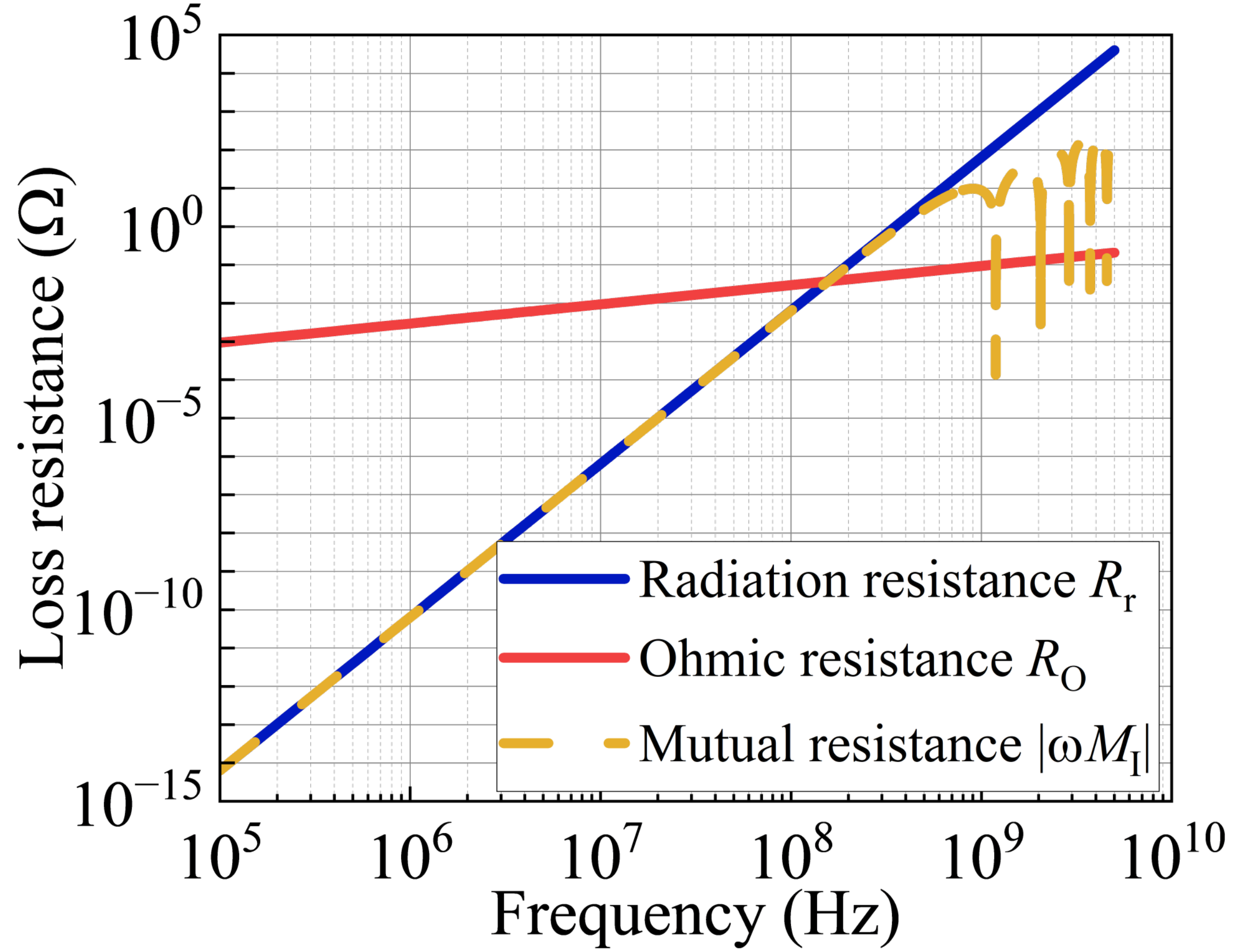}
	\caption{Comparison of Ohmic resistance \eqref{eq:18} and radiation resistances of one loop antenna \eqref{eq:19}, together with the mutual resistance between two antennas $\omega M_{\rm I}$. The loop radius $a=36 $~mm, and $d=5a$.}
	\label{fig:MD_losses}
\end{figure}

To understand these effects, we need to consider the relative values of the Ohmic resistance and the radiation resistance, because the significance of radiation loss suppression is determined by the relative level of losses due to dissipation in the antennas and parasitic radiation into far zone. To estimate this ratio, we plot in Fig.~\ref{fig:MD_losses} all relevant resistances as functions of the frequency for the considered example of two loops.

We see that at low frequencies the loss resistance strongly dominates over the radiation resistance. For this reason, radiation suppression has negligible effect on performance in the quasi-static regime. The power transfer efficiency is low because the quasi-static coupling between loops at such large distance is weak. On the other hand, we observe that exactly in the frequency range of the radiation loss suppression for each considered value of the loss resistance, the power transfer efficiency $PTE$ has a strong peak. 

For electrically small transfer distances, the imaginary part of the mutual inductance $M_{\rm I}$ is negligibly small, and 
the general formula (\ref{eq:10}) tells that  
higher efficiency corresponds to smaller distances $d$ between the antennas (to larger values of the mutual inductance $M_{\rm R}$). It is the case of conventional inductive WPT devices. Now we see that at electrically larger distances between the two antennas there is another possibility to realize high-efficiency wireless power transfer. This regime takes place when $M_{\rm I} $ is large and the far-field radiation is suppressed. In this regime, we do not need high values of the real part of the mutual inductance. 
These results open a possibility to realize high-efficiency wireless power transfer to distances that are large compared to the antenna sizes. 

\begin{figure*}
	\centering
	\includegraphics[width=0.95\textwidth]{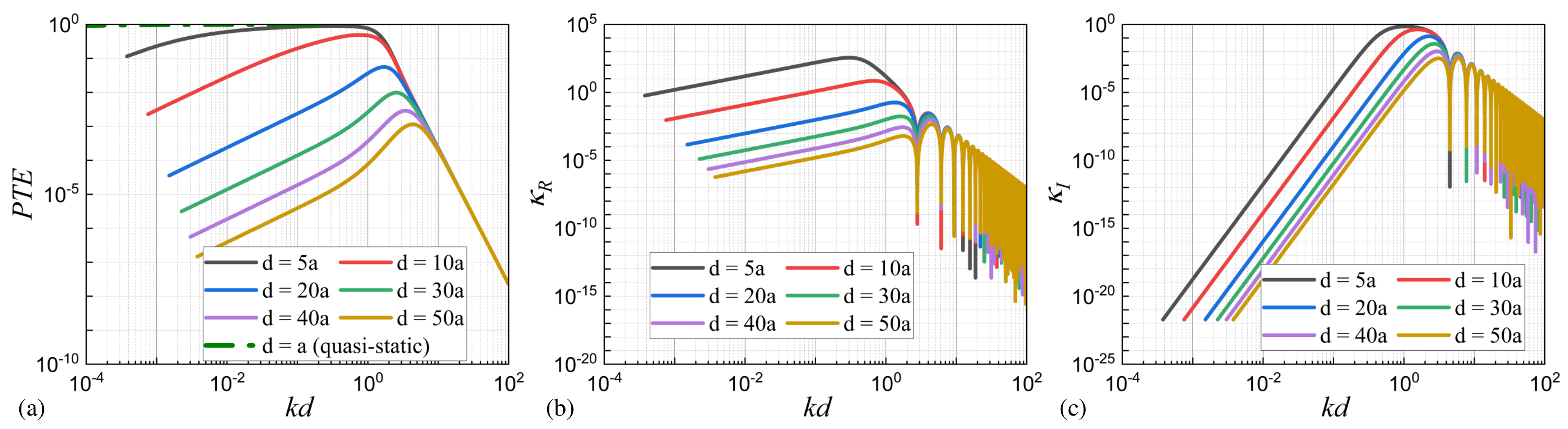}
	\caption{Coaxial arrangement: a) $PTE$ versus $kd$; b) $\kappa_{\rm R}$ versus $kd$; c) $\kappa_{\rm I}$ versus $kd$ for different values $d/a$ from $5$ to $50$. 
	} 
	\label{fig:Kappa_coaxial}
\end{figure*}

\begin{figure*}
	\centering
	\includegraphics[width=0.95\textwidth]{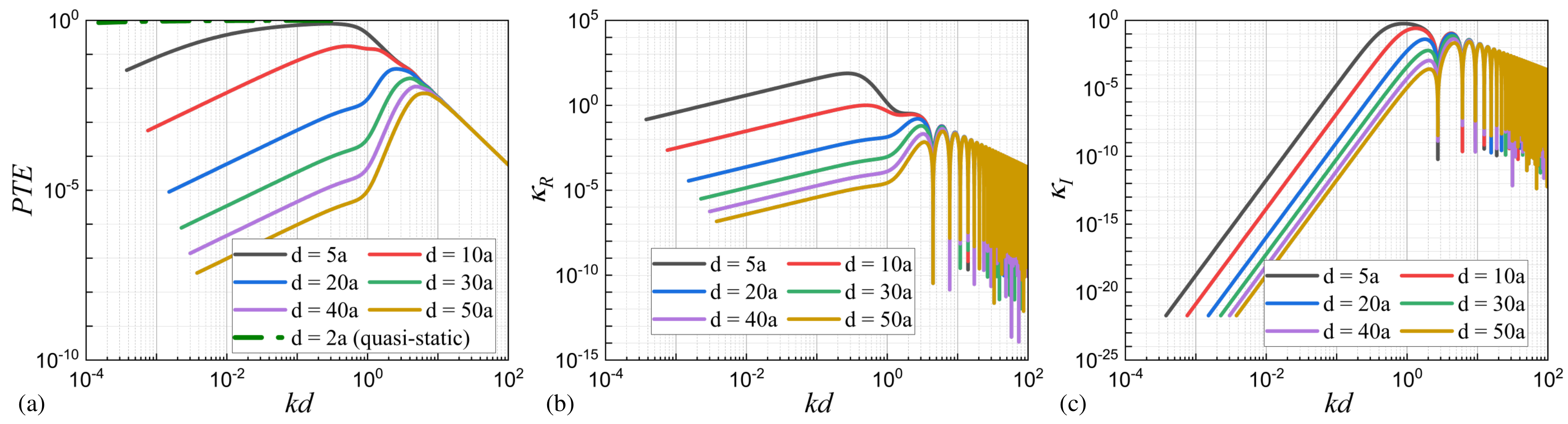}
	\caption{Coplanar arrangement: a) $PTE$ versus $kd$; b) $\kappa_{\rm R}$ versus $kd$; c) $\kappa_{\rm I}$ versus $kd$ for different values $d/a$ from $5$ to $50$.  
	} 
	\label{fig:Kappa_coplanar}
\end{figure*}


\subsection{Analytical Results for Optimized Mid-range WPT}

Next, we study high-frequency mid-range power transfer in more detail, considering various transfer distances and both coaxial and coplanar orientations of the loops. In our analysis, we focus on the almost radiation suppression regime in the mid-range  WPT, where the corresponding distance $d$ is not very small compared to the wavelength but large as compared to the loop radii $a,b$. In this case  $M_{\rm I}$ is large  in a rather broad range of frequencies, which helps to achieve high power transfer efficiency. 

In order to better clarify the roles of $M_{\rm R}$ and $M_{\rm I}$ in attaining the maximal $PTE$, we introduce two normalized coupling parameters
$\kappa_{\rm R}=\omega^2M_{\rm R}^2/R_1R_2$ and $\kappa_{\rm I}=\omega^2M_{\rm I}^2/R_1R_2$, 
so that Eq.~\eqref{eq:10} takes the form
\begin{align}
PTE=1-\dfrac{2}{1+\sqrt{\dfrac{1+\kappa_{\rm R}}{1-  \kappa_{\rm I}}}}. 
\label{eq:10kappa} 
\end{align}
We fix the radii of the loops $a=b=36$~mm and plot their values versus the normalized distance $kd$ for many values 
of $d$ from $d=5a$ to $d=50a$ and compare these coupling parameters with the similar plots of $PTE$. For each curve, the value of $d$ is fixed, that is, the curves effectively show dependence on $k$ (the frequency).
The three corresponding plots are presented in Fig.~\ref{fig:Kappa_coaxial} (coaxial arrangement) and  
Fig.~\ref{fig:Kappa_coplanar} (coplanar arrangement). 

We see that at $kd<1$, at all mid-range distances $PTE$ is low because the real part of the mutual inductance is large only  in the near-field zone, when $d<a$,  and also the imaginary part $M_{\rm I}$ is small compared to the Ohmic resistance. In other words, when $d\gg a$ and $kd < 1$, the loops are weakly coupled, leading to small values of $PTE$. In the conventional quasi-static regime of WPT, when $kd\ll 1$ and the antenna size $a$ is comparable to the distance $d$, the mutual reactance $\omega M\approx \omega M_{\rm R}$ is significantly larger than the Ohmic resistance. Therefore, in this conventional case of small distances and large loops we can achieve a high $PTE$ according to~(\ref{eq:10}) and \eqref{eq:10kappa}. Since the dipole-moment model of loops is not applicable in the quasi-static case, we have studied the quasi-static regime numerically and calculated the $PTE$ in both coaxial and coplanar arrangements. The results are shown as dash-dot lines in Fig.~\ref{fig:Kappa_coaxial}(a) and Fig.~\ref{fig:Kappa_coplanar}(a), respectively.
For very large electromagnetic distances  where $kd\gg 1$, both coupling parameters and, respectively, $PTE$ again becomes low. This is because the mutual impedance decreases in the far zone, whereas the radiation resistance of antennas rapidly grows versus frequency, i.e., versus $kd$. Thus, we can conclude that at the mid-range distances, where $kd\sim 1$, for any $d\gg a$ there is a range of optimal electrical distances $kd$ 
corresponding to the maximal total coupling and, therefore, to  the maximized $PTE$.

For the coaxial case, when there is no contribution of far fields to the power transfer, the decrease of the $PTE$ at high frequencies is more rapid than its increase at low frequencies. For the coplanar case, the $PTE$ curves are more symmetric, because at high frequencies not only the radiation loss increases, but also the far-field coupling becomes stronger.  

For the coaxial arrangement, the maxima of $\kappa_{\rm R}$ and $\kappa_{\rm I}$ as functions of $kd$ occur at different values of $kd$. 
For example, for $d=5a$ these maxima take place at $kd\approx 0.3$ and $kd\approx 0.5$, respectively. 
For $kd=0.5$, we have from~(\ref{eq:15}) and~(\ref{eq:19})
\begin{equation}
\omega M_{\rm I}={\pi\eta k^2a^4\over 2d^2}\left({\sin kd\over kd}-\cos kd
\right)\approx {\pi\eta k^4a^4\over 6.15}, 
\end{equation} 
which is close to $ {\pi\eta k^4a^4\over 6}= R_{\rm r1,r2}$. This shows that the radiation resistance of each of the two loops is almost compensated by the mutual resistance, and this case corresponds to 
the optimal non-radiative power transfer. Due to some residual radiation and the presence of losses  $PTE|_{kd=0.5}\approx 0.89$, i.e., about 11\% of power is dissipated in resistors $R_{\rm O1,O2}$ and radiated. 
However, the maximum of $PTE$ (also about $0.9$) holds not at $kd=0.5$ but at $kd=0.4$. At this point 
the sum $\kappa_{\rm R}+\kappa_{\rm I}=\omega^2|M|^2/R_{1}R_2$ is maximal. 
This regime can be still called the radiation-suppression  regime because the maximum of the function $\kappa_{\rm I}(kd)$, as we can see in  
Fig.~\ref{fig:Kappa_coaxial}, is wide and overlaps with the maximum of $\kappa_{\rm R}(kd)$, which is also wide in accordance to 
Fig.~\ref{fig:Kappa_coplanar}. We note that for our specific example $kd=0.4$ in the case $d=5a$ corresponds to the frequency of 110~MHz.
Below we will see that this is the optimal WPT frequency for the considered example of two loops. 

In the coplanar arrangement, maxima of $\kappa_{\rm R}(kd)$ and $\kappa_{\rm I}(kd)$ occur at $kd=0.4$ and $kd=0.8$, respectively, whereas $\kappa_{\rm R}|_{kd=0.4}\approx 190$ and $\kappa_{\rm I}|_{kd=0.8}\approx 0.7$. In this case the $PTE$ also attains its maximum close to $0.7$ at $kd=0.6$ in the middle between the maxima of $\kappa_{\rm R}(kd)$ and $\kappa_{\rm I}(kd)$.     

In the coaxial case, the far field plays a parasitic role, because the Tx loop does not radiate into the Rx direction. However, the important coupling parameter $\kappa_{\rm I}$ attains the maximum at a substantial electromagnetic distance ($kd >1$) because the expression for $M_{\rm I}$ comprises two terms that  compete with one another when $kd$ is not very small and not very large. At still larger distances the efficiency quickly decays, since far-field coupling becomes small. For example, the maximal $PTE$  at $d=50a$ is as small as $0.001$.  

In the coplanar mid-range ($1 < kd < 10$) case, the far field brings a contribution into both $M_{\rm R}$ and $M_{\rm I}$ but also brings radiation loss measured by $R_{\rm r1,r2}$. The optimal WPT is achieved for a substantial electromagnetic distance $kd\approx 5$ where $\kappa_{\rm I}$ is maximal and $\kappa_{\rm R}$ is close to the middle-range maximum. The far field plays an advantageous role at this distance, and $PTE$ in this optimal case is equal to $0.007$. In other words, due to the contribution of radiation the maximal $PTE$ in the mid-range region turns out to be 7 times higher than that we have obtained for the same large physical distance in the coaxial arrangement. In terms of the electromagnetic distance $kd$, the advantage granted by the far field is even higher because the optimal $kd$ for the coplanar case ($kd=5$) is larger than that for the coaxial case ($kd=4$).

To conclude the discussion of Figs.~\ref{fig:Kappa_coaxial} and \ref{fig:Kappa_coplanar}, it is worth noting that in the far zone ($kd>\pi$) there are periodic maxima and minima of $\kappa_{\rm R,I}(kd)$, whereas maxima of $\kappa_{\rm R}(kd)$ coincide with the  minima of $\kappa_{\rm I}(kd)$ and vice versa (that results in a smooth decrease of $PTE$ with the normalized distance $kd$). Two adjacent maxima (and minima) of $\kappa_{\rm R}$ (and $\kappa_{\rm I}$) are distanced by $2\pi$ because the phase shift between the  Tx excitation  and the field at the Rx antenna position oscillates with the period $\lambda$. At such large distances, we approach the far-field WPT regime, where one needs to use directive antennas instead of small dipole antennas.

\begin{figure*}
	\centering
	\includegraphics[width=0.85\textwidth]{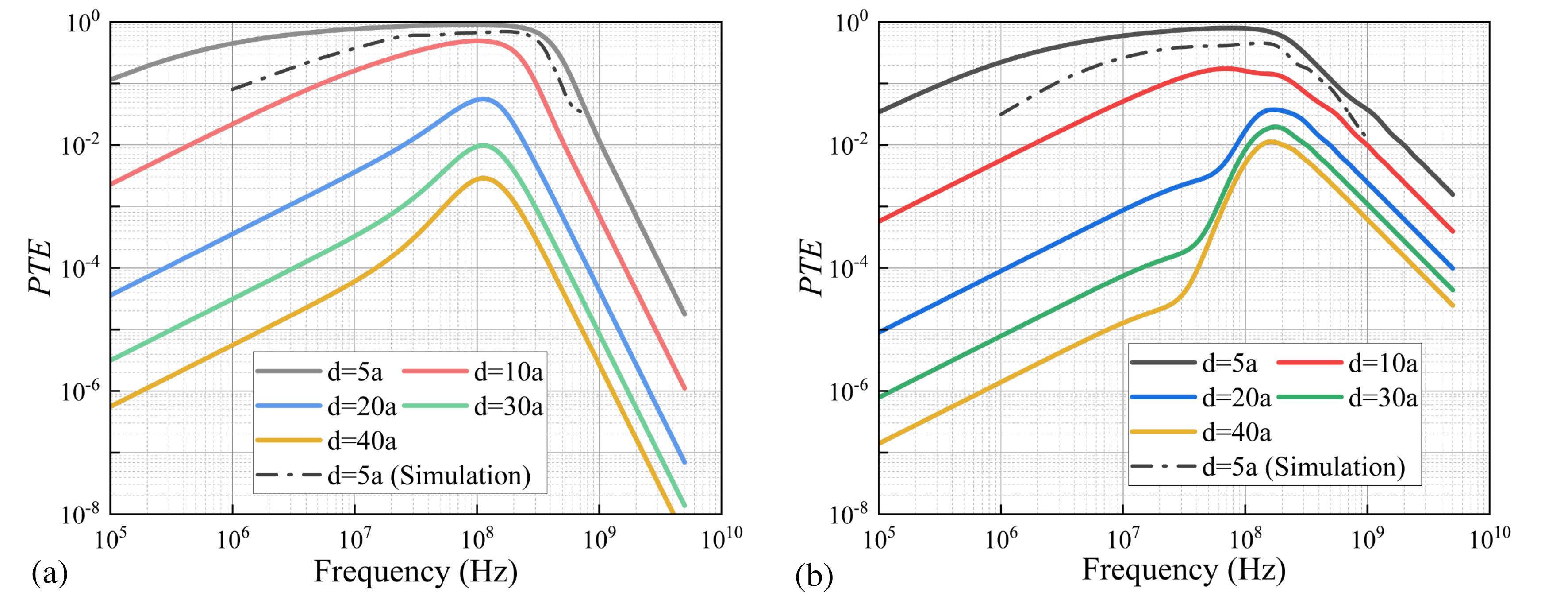}
	\caption{$PTE$ in a broad frequency range for different normalized transfer distances $d/a$.\\ a) coaxial arrangement; b) coplanar arrangement. 
	} 
	\label{fig:PTE_distance}
\end{figure*}

\begin{figure*}
	\centering
	\includegraphics[width=0.85\textwidth]{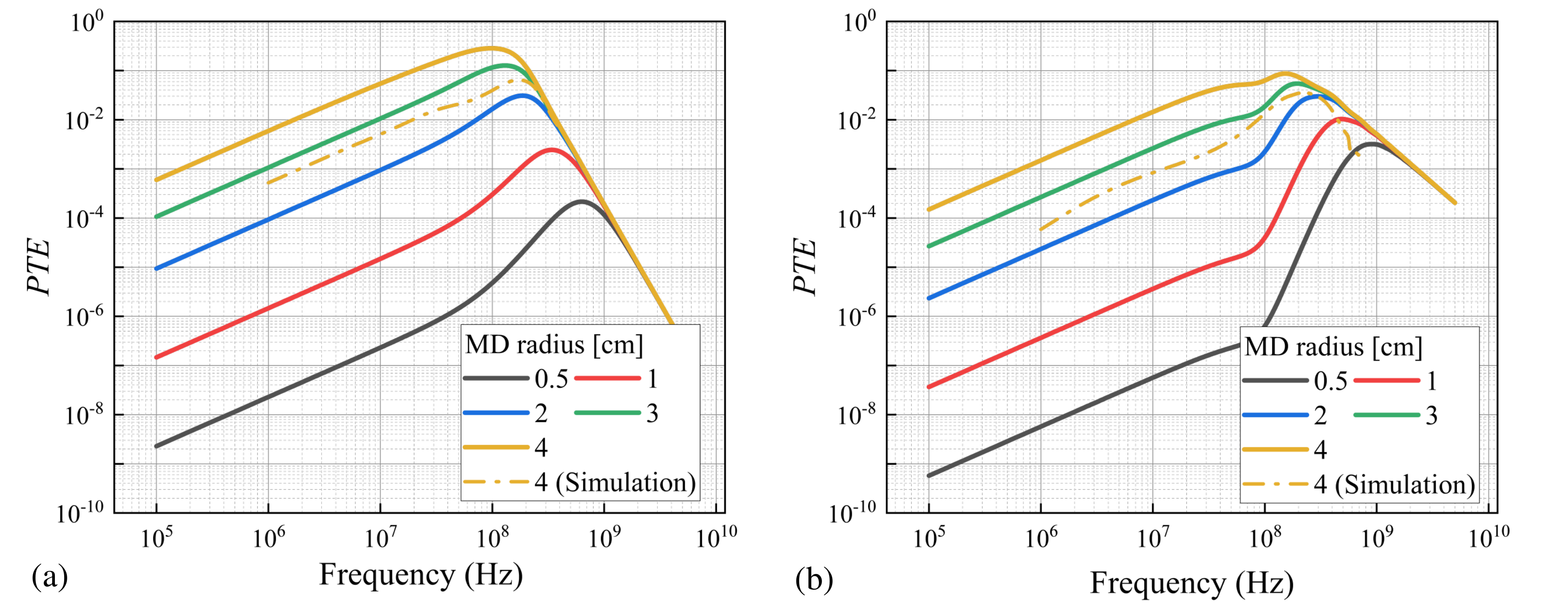}
	\caption{$PTE$ in a broad frequency range for different antenna sizes. The distance between the two antennas $d=0.5$~m. a) Coaxial arrangement; b) coplanar arrangement. 
	} 
	\label{fig:PTE_dimension}
\end{figure*}

Figure~\ref{fig:PTE_distance}(a) shows  ${PTE}$ for the coaxial arrangement at frequencies from 500~kHz to 5~GHz with the normalized distance $d/a$ changing from 5 ($d=18$~cm)  to 50 ($d=1.8$~m). These results allow us to find the optimal operation frequency for practical long-distance (transfer distance much larger than the antenna sizes) WPT devices. Therefore, we performed a numerical validation of the model using a full-wave electromagnetic solver CST Studio 
together with the  ADS simulation tool that modeled lossless matching circuits for the receiving loop at any frequency. 

In Fig.~\ref{fig:PTE_distance}(a) besides of the analytical results (solid curves) we show the simulated $PTE$ for $d=5a$ (dash-dotted curve). Qualitatively, this curve confirms the analytical model. Quantitative differences are as follows: the optimal operation frequency 160 MHz instead of 110 MHz, and $PTE_{\rm max}=0.7$ instead of $0.9$. The reasons of the disagreement are clear. In the analytical model, we replace the Tx loop by a magnetic dipole, whereas in the simulations it is a split loop fed by a generator connected to a finite gap of width $g$. Therefore, besides of a magnetic current mode, there is an electric dipole mode induced in the Tx loop. The electric coupling of two loops results in the electric dipole mode also in the Rx loop. Parasitic impact of electric coupling leads to a shift of the optimal frequency and makes the cancellation of radiation less effective. At a first glance, this parasitic effect can be made negligible decreasing $g$, however, simulations show that $g\ll 1$~mm correspond to a so large split capacitance that it completely shunts the loop preventing its excitation. 

In principle, it is possible to analytically optimize the loop even if the split is large. For such optimization it is necessary to take into account 
the electric coupling of loops together with the magnetic one. However, it is not the  purpose of this study. We aim to developing a theory clearly explaining the role of near and far fields in mid-range WPT, which calls for a simplest possible analytical model to be used. 
Therefore, in simulations we simply numerically optimized the gap width $g$. The best correspondence of simulations and analytical theory holds when $g=1$ mm and when the wire of the loop is slightly enlarged in the near vicinity of the split edges. Note that the problem of a finite gap  
exists only for the Tx loop and does not arise for the Rx loop. The electromagnetic solver allows us to load it by a lumped resistance whose value is determined by formula~(\ref{eq:9}).  

The corresponding numerical curve for the coaxial arrangement and $d=5a$ is depicted in Fig.~\ref{fig:PTE_distance}(a). In Fig.~\ref{fig:PTE_distance}(b),
we present a similar set of theoretical curves and the validation of the upper curve ($d=5a$) for the coplanar arrangement.   

Besides of comparison of the analytical results with full-wave simulations, Figs.~\ref{fig:PTE_distance}(a) and (b) lead to the following observations.
First, we see that the optimal frequency of 110~MHz is the same for the coaxial arrangement at any considered $d/a$. The coincidence of these frequencies for the 
case $d=5a$ and $d=50a$  has been already discussed above. Now, we observe it in the whole range $d/a$.   
Second, for the coplanar arrangement, as we can see in Fig.~\ref{fig:PTE_distance}(b), the absolute maximum of $PTE$ shifts versus $d/a$. 
In accordance to the theory, it shifts from about 80 MHz ($d=5a$, $PTE_{\rm max}=0.5$) to nearly 180~MHz ($d=50a$, $PTE_{\rm max}=0.007$).
Third, for both arrangements the optimal frequency range is about 100--200 MHz, that is significantly lower than the self-resonance frequency of a wire loop with the radius $a=36$~mm. This frequency range results from the formulas of the analytical model, confirmed by numerical simulations, and it is general: if a WPT device is based on two MDs, the maximal efficiency is achieved at a frequency much lower than the self-resonance of the individual antenna. However, this frequency is higher than those corresponding to the quasi-static regime -- the regime in which $\omega M_{\rm I}\ll R_{\rm r1,r2}$.   

In \cite{Skrivernik}, numerical calculations similar to those presented in Fig.~\ref{fig:PTE_distance} were done for a different type of WPT. In \cite{Skrivernik}, a spherical muscle phantom with a very small receiving antenna in the center was excited by a uniformly distributed source over the sphere surface. That ideal Tx was  creating a fixed incident electromagnetic power at all frequencies (reflections from the sphere boundary were assumed to be eliminated by an ideal antirefelecting coating). The Rx antenna was assumed to be perfectly matched with the load and $R_{\rm L}$ was optimized for the maximal power transfer at every frequency. Full-wave simulations were performed in \cite{Skrivernik} 
with the aim to find the optimal frequency of WPT for both electric and magnetic Rx antennas. Interestingly, the set of curves plotted in Fig.~3 of \cite{Skrivernik} for different $d/a$ qualitatively resembles our plots in both Figs.~\ref{fig:PTE_distance}(a) and (b). 
The main qualitative difference is a negative shift of the optimal frequency versus $d$ in \cite{Skrivernik}. 
This shift definitely results from electromagnetic energy dissipation in the medium. In the muscle tissue, the near-field coupling exponentially decays when $d$ increases, and this decay factor is proportional to the frequency. Meanwhile, in the considered case two MDs are located in free space, and their near field at a given distance weakly depends on the frequency. However, the main qualitative result of the  study \cite{Skrivernik} is the optimal frequency range for WPT located below the self-resonance of Rx but higher than the 
low-frequency region in which the quasi-static approximation for the whole system is applicable. This result fits our one. In work \cite{Shamonina1}, the 
coupling of two MD-based antennas in a lossy medium is considered with the purpose to optimize the antenna sizes for the given operation frequency. Again, the operation frequency turns out to be located between the low-frequency region and the band of the antenna self-resonance. And the same refers to the range of optimal operation frequencies when the Tx and Rx antennas are located in free space.   

In Figs.~\ref{fig:PTE_dimension}(a) and (b), we present $PTE$ versus frequency for both cases of coaxial and coplanar arrangements for a comparatively large fixed transfer distance $d=0.5$~m and different loop radii $a$. We see that the optimal frequency decreases versus $a$. This decrease, evidently, results from the increase in the coupling for given frequency when the loop gets larger. The top theoretical curves in Fig.~\ref{fig:PTE_dimension}(a) and (b), i.e., the case $a=4$~cm, is validated by numerical simulations. Again, we note a qualitative agreement, but the simulated $PTE_{\rm max}$ is triply lower than the theoretical prediction for the coaxial arrangement and twice lower for the coplanar one. In other words, the numerical disagreement between the theory and simulations in the case $d=12.5a$ is much larger than that in the case $d=5a$, when the radiation suppression regime is achieved. This result confirms our insight that the main inaccuracy of the magnetic dipole model is not in the approximate formulas for the mutual inductance, but mostly in negligence of the electric dipole mode. For the loop $a=4$~cm this mode is more noticeable than for the loop $a=3.6$~cm operating at the same frequency.     

To conclude this section, let us formulate the most important theoretical observations: 
\begin{itemize}
\item In the mid-range coupling regime between two electrically small loops there is a possibility to reach reasonably high power transfer efficiency due to suppression of radiation into far zone (compensation of the radiation resistances of both loops);
\item This regime is realized by tuning the load impedance to the optimal value for the maximal transfer efficiency;
\item 
The optimal frequency for WPT between two small loops to distances much larger than the loop sizes in both cases of the arrangement of Tx and Rx loops lies above the low frequency region (in which the coupling of two antennas is quasi-static and strong) but below the resonance band of the individual loop antennas;  

\item 
To realize effective wireless power delivery at the maximally possible distance between small antennas, the mid-range regime with the coplanar arrangement of loops is suitable. In this case the far-field coupling grants a significant  improvement of $PTE$ compared to the case when WPT holds solely due to near fields, which makes the coplanar arrangement of loops advantageous compared to the coaxial one. 
\end{itemize}

\begin{figure*}[t]
	\centering
	\includegraphics[width=\textwidth]{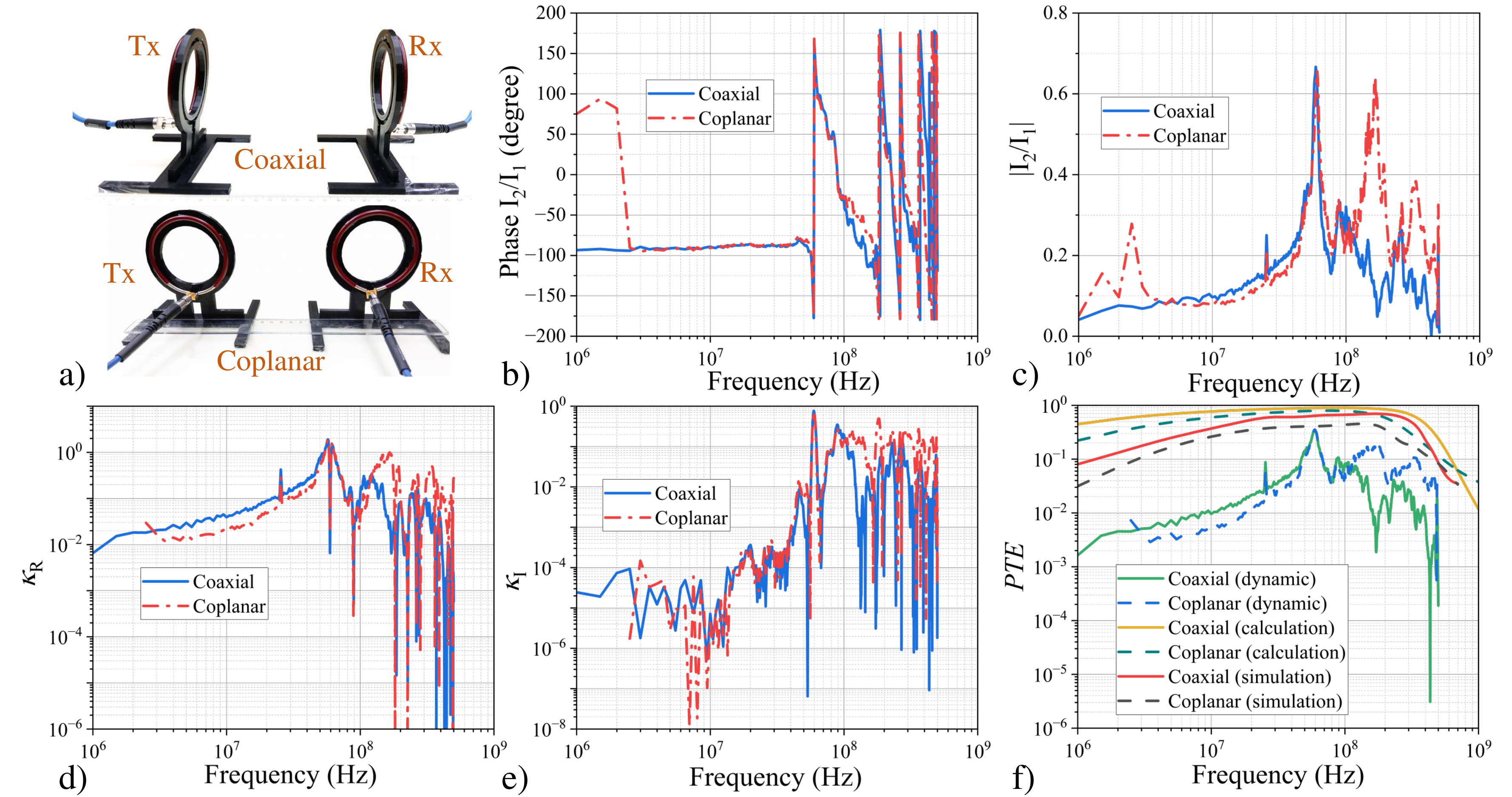}
	\caption{Experimental setup and results: a) Two loop antennas in coaxial and coplanar arrangements; b) Phase difference of the loop currents; c) Magnitude ratio $I_2/I_1$; d) $\kappa_{\rm R}$; e) $\kappa_{\rm I}$; f) $PTE$ comparison between calculated, simulated, and measured results for coaxial and coplanar antenna arrangements.
	} 
	\label{fig:Experiment}
\end{figure*}

\section{Experimental Verification}\label{sec:Experiment}

In this section, we present an experimental validation of the theoretical results. The experimental setup consists of two loop antennas designed for the frequency range of interest. In order to minimize the effect of the electric-dipole mode, we use shielded loop antennas made from  coaxial cables. The center conductor of the cables forms transmitting and receiving loops, and the cable screen serves as a shield, minimizing the electric-dipole mode current. The two terminals of the inner copper wire of the loop are welded to the center pins and the ground parts of SMA connectors. The shield is cut close to the connection points. Both Tx and Rx loops have the same radius of 3.6~cm and are placed at a center-to-center distance of 18~cm ($d=5a$) for coaxial and coplanar arrangements, as shown in Fig.~\ref{fig:Experiment}(a). For each arrangement, the antennas are connected  to two ports of the ZND Vector Network Analyzer (VNA) to measure $Z$-parameters of the system in a wide frequency range, which allows evaluating the optimal frequency and $PTE$ for the optimal load. The measured $Z$-parameters are used for analytical calculations of the optimal load impedance and power transfer efficiency using the formulas presented above. The experimental results are shown  in Fig.~\ref{fig:Experiment}.

We see that the phase difference and magnitude  ratio  $I_2/I_1$ of the currents in the transmitting and receiving loops show a qualitative agreement with numerical calculations (see blue dashed lines in Fig.~\ref{fig:rad_lossy}). At the optimal frequency of 59.38~MHz for the coaxial and coplanar cases, the currents in the two loops have approximately opposite phases ($-177.451^{\circ}$/$-175.341^{\circ}$) and similar amplitudes (0.67/0.607), as is seen in Figs.~\ref{fig:Experiment}(b)-(c). Here, the power transfer efficiency  ($PTE$) has a peak (0.35/0.353) corresponding to the peak of $\kappa_{\rm I}$, although $\kappa_{\rm R}$ is small, as is seen in Figs.~\ref{fig:Experiment}(d)-(f). 

Let us also  compare  $PTE$ in the dynamic regime with the system working in the conventional quasi-static coupling regime. In this case, the loop coupling is defined by the mutual inductance, conventionally determined by the geometry and positions of the loops. To do that, we measured the mutual inductance and the $Z$-parameters of the two coupled loops at distance $d=5a$ at the frequency of 4~MHz.  The measured mutual inductance is approximately 0.56 nH for the coaxial case (that is around 0.508 nH in numerical calculations) and 0.384 nH for the coplanar case (0.313 nH in numerical calculations). 
Substituting the measured mutual inductances and measured resistances $R_{1,2}={\rm Re}(Z_{11,22})$ into formula \eqref{eq:10}, we find an estimation for the maximal achievable $PTE$ at this distance $PTE\approx 0.0065$ for the coaxial arrangement and $PTE\approx 0.0035$ for the coplanar arrangement.  These values are dramatically smaller than those corresponding to the theoretical, simulated, and experimental curves shown in Fig.~\ref{fig:Experiment}(f), confirming a possibility to significantly enhance power transfer efficiency using properly tuned dynamic coupling. 

Overall,  we observe qualitative agreement between the theoretical and experimental results. Quantitative differences are a small shift in the optimal frequency range and an order of smaller  magnitude of the $PTE$ in the experiment compared to numerical calculations. This big difference can be explained by losses in the dielectric support and in the galvanic contacts.

\section{Conclusion}

In this paper, we have developed the dynamic theory of wireless power transfer between two antennas, specializing it for identical loop antennas. The theory covers a very broad range of frequencies and refers to both short-range and mid-range distances between the loops. 
Assuming that the Rx antenna reactance is properly matched at every possible operational frequency and its useful load resistance is similarly optimized, we aimed to find the optimal frequency range for a given loop size and track how and why the power transfer efficiency changes with the distance. We studied two most important mutual arrangements of the loops: the coaxial and the coplanar ones. In the first arrangement, the coupling between the Tx and Rx is solely due to the near fields, in the second case the near-field coupling is low and the radiative, far-zone coupling exists.  

We have shown that when the distance $d$ is very small as compared to the antenna size, and dissipation is negligible, the power transfer efficiency can approach unity at whatever (enough low) operation frequency, and the only issue is proper matching of antennas. However, if $d$ is larger than the loop diameter $2a$, finding the optimal operation frequency becomes crucial for WPT system design. For the coaxial arrangement, we have found the optimal frequency which turned out to be unique for a broad 
range of distances ($5a \le d\le 50a$). For the coplanar arrangement, the optimal frequency depends on the distance, but this dependence is quite weak. In both cases, the optimal frequency lies below the self-resonance of the loop but above the band in which the quasi-static model is applicable for the WPT system. In this frequency range the interaction between the loops is dynamic, measured with both inductive and resistive parts of the mutual impedance. 

The maximum value of the $PTE$ turns out to be quite high in spite of the geometrically and  electromagnetically substantial distance between the two antennas. In this optimal regime,  the mutual coupling of two loops suppresses the radiation from the system, and almost all the power, which is not lost due to parasitic dissipation in the Tx antenna, is transferred to the Rx. 
We have thoroughly studied the prerequisites and peculiarities of this regime, that holds for both coaxial and coplanar arrangements, considering  contributions and advantages of the near-field and far-field couplings for mid-range distances.
The developed analytical model has been verified by full-wave simulations and partially validated experimentally. We believe that the regime of the radiation suppression is very important not only because it grants high $PTE$ for substantial transfer distances but also because it prevents parasitic heating of the ambient around the WPT system. 

\section*{Acknowledgements}
This work was partially supported by the Academy of Finland, project 338786 and the Academy of Finland postdoctoral researcher grant 333479. 



\end{document}